\documentclass[journal,twoside,web]{ieeecolor}
\usepackage{generic}
\usepackage{cite}
\usepackage{amsmath,amssymb,amsfonts}
\usepackage{graphicx}
\usepackage{algorithm,algorithmic}
\usepackage{hyperref}
\hypersetup{hidelinks=true}
\usepackage{textcomp}
\usepackage{multirow}
\usepackage{makecell}
\newtheorem{lemma}{Lemma}
\usepackage{tikz}
\newcommand\submittedtext{%
  \footnotesize This work has been submitted to the IEEE for possible publication. Copyright may be transferred without notice, after which this version may no longer be accessible.}

\newcommand\submittednotice{%
\begin{tikzpicture}[remember picture,overlay]
\node[anchor=south,yshift=10pt] at (current page.south) {\fbox{\parbox{\dimexpr0.65\textwidth-\fboxsep-\fboxrule\relax}{\submittedtext}}};
\end{tikzpicture}%
}

\def\BibTeX{{\rm B\kern-.05em{\sc i\kern-.025em b}\kern-.08em
    T\kern-.1667em\lower.7ex\hbox{E}\kern-.125emX}}
    
\renewcommand{\arraystretch}{1.5}

\usepackage{array, booktabs}
\newcolumntype{B}{!{\vrule width 1.4pt}}

\begin{document}
\title{Untangling the Geometry and Speed for RF Sensing Spectrograms}
\author{Mert Torun, \IEEEmembership{Member, IEEE}, Darius Cuenca, \IEEEmembership{Member, IEEE}, and Yasamin Mostofi, \IEEEmembership{Fellow, IEEE} \vspace{-18pt}
\thanks{This work was supported in part by NSF CNS award 2226255, and in part by ONR award N00014-23-1-2715.}
\thanks{Mert Torun, Darius Cuenca, and Yasamin Mostofi are with the Department of Electrical and Computer Engineering, University of California, Santa Barbara (e-mail: merttorun@ucsb.edu; dariuscuenca@ucsb.edu; ymostofi@ece.ucsb.edu).}
\thanks{Approval of all ethical and experimental procedures and protocols was granted by the Institutional Review Board (IRB) Committee at UCSB.
}
}

\maketitle
\submittednotice

\begin{abstract}
A fundamental challenge in RF sensing is that Doppler signatures observed by a link entangle the target’s motion with the sensing geometry, resulting in limited applicability to unconstrained real-world settings. In this paper, we establish a new foundation for physically interpretable RF sensing that disentangles reflector speed from geometry, jointly recovering the speed, geometry factor, relative amplitude, and width of each dominant Doppler ridge. More specifically, we first develop a compact parametric representation of WiFi spectrograms and establish its low-dimensional structure through a systematic computer-vision analysis of a large and diverse human-activity dataset, thereby providing a tractable foundation for learning. Building on this representation, we then design a physics-informed autoencoder whose structured bottleneck and differentiable RF forward model enforce physically meaningful estimates of reflector speed and geometry. We further introduce a synthetic-to-real training framework, eliminating the need for real WiFi training data. We extensively validate the proposed framework under both known and time-varying geometries, using both independently generated synthetic test sets and 31 real WiFi experiments. The results demonstrate the superior performance in speed and geometry extraction, robustly recovering the underlying geometry, speeds, Doppler-ridge amplitudes, and ridge widths across all settings, while substantially outperforming the strongest baselines.
\end{abstract}

\begin{IEEEkeywords}
Speed extraction, geometry inference, micro-Doppler analysis, WiFi sensing, human-centered computing, RF sensing, environmental sensing, smart spaces.  
\end{IEEEkeywords}

\section{Introduction} \label{sec:Introduction}

\IEEEPARstart{I}{n} recent years, the widespread presence of wireless signals has established Radio Frequency (RF) based sensing as a powerful technique for non-invasive sensing, perception, and activity monitoring. By analyzing the transient properties of traveling wireless waves, in particular the micro-Doppler traces, researchers have tackled diverse challenges, such as human activity recognition~\cite{sensefi, widar3, torun2023wi}, smart health \cite{ei_har, korany2021nocturnal, parsay2025gait}, imaging~\cite{pallaprolu2022wiffract}, and crowd semantics~\cite{choi2022wi,pallaprolu2024crowd,korany2021counting}, among others. Historically, vision-based systems and cameras have been the primary tool for such analytics. However, growing privacy concerns, lighting-limited environments, and occlusion limitations have since pushed researchers toward RF-based approaches. This has driven a surge in both classical signal processing techniques and machine learning algorithms designed to interpret RF signals for a variety of applications.

Despite these advances, a fundamental limitation remains largely unresolved: the Doppler signatures observed by a single RF link entangle the target’s true motion with the sensing geometry. As a result, most existing RF sensing systems avoid this ambiguity by assuming favorable/fixed motion directions, relying on known geometry, or treating geometry-induced variation as a nuisance to be suppressed. More specifically, when dealing with moving subjects, existing RF sensing work has frequently circumvented spatial variation by strictly controlling the target's geometry. In these applications, for instance, directional assumptions are hard-coded into the experimental design, requiring individuals to face specific directions or move along fixed, predetermined, and favorable paths relative to the transceiver \cite{lv2017device, zhang2022metaganfi, wang2019wipin}. Such artificial constraints then let models map signal features directly to kinematics. However, such assumptions substantially limit deployment in natural, unrestricted settings, where people may move along arbitrary paths and where both the motion speed profile and the underlying geometry are unknown. This shortcoming is further highlighted when extending RF sensing to non-human targets, such as tracking quadrupedal gaits for veterinary care or wildlife monitoring, where targets of interest clearly cannot be instructed to align their movements. This motivates the central problem of this paper: recovering physically meaningful reflector profiles (e.g., speed, amplitude, etc.) from a single WiFi spectrogram while simultaneously estimating the geometry factor that shaped the observed Doppler signatures.

Another line of existing work addresses spatial variation indirectly by treating changes in target position, orientation, or environment as a domain-shift problem. For instance, past work attempted to use domain adaptation and one-shot calibration to address geometry impacts~\cite{zhang2018crosssense, one_shot_har, zhang2022metaganfi}. For instance, ~\cite{ei_har, datta_2026, gesfi_2025} proposed that adversarial learning and test-time adaptation can be used for domain-invariant feature selection. While valuable effort has been made towards generalizability to unseen domains, these methods aim to make internal representations less sensitive to domain shifts but cannot address the fundamental problem of untangling geometry and speed profile. More specifically, these methods treat the sensing geometry as a nuisance to be suppressed and cannot recover both the underlying geometry and physical speeds.  

Finally, another line of existing work addresses the performance degradation caused by geometry impact through adding additional resources. For instance, \cite{wang_pulse_doppler} uses two vertically separated radars, at foot and torso levels. \cite{wu2020gaitway} identifies stable walking periods to recover bulk gait speed and stride features. \cite{widar3} combines Doppler signatures from multiple WiFi links to reconstruct a body speed profile while requiring external position and orientation information. A more recent study, \cite{gao2026speedfi}, estimates a continuous torso speed profile by fusing two orthogonal WiFi links. Similarly, \cite{rf_pose} recovers body skeleton utilizing a dense antenna array. In summary, these methods ease the geometry restrictions. However, they require more resources or specialized hardware, lack scalability, or demand cumbersome data collection that is not readily available. In summary, prior work either enforces ideal motion directions, assumes knowledge of motion directions, adds spatial diversity through multiple WiFi links or antenna arrays, or aims to construct domain-invariant features for a specific task. 

To the best of our knowledge, none of the existing work jointly estimates major reflector speed profiles while simultaneously explaining the underlying geometry factor for a single WiFi link. The gap we address is therefore not the absence of speed estimation or cross-domain generalization, but the absence of a model whose explicit output physically factorizes a single-link Doppler observation into its speed and geometry components. In this work, we propose a contextual estimation, in which feasible geometry-component speed combinations explain the underlying physical parameters that create a WiFi spectrogram. More specifically, we design a physics-informed autoencoder pipeline that explicitly decouples spatial geometry from the target's kinematics. We implement this principle through a fully differentiable, physics-informed autoencoder whose bottleneck jointly estimates reflector speeds with Fourier series coefficients and the trajectory parameters that determine the geometry factor. An analytical RF decoder then reconstructs the spectrogram from the predicted speeds, geometry factor, Doppler ridge amplitudes, and widths, requiring the bottleneck variables to remain physically meaningful. Equally important, our pipeline not only untangles and infers both the geometry and speed profile, but can also estimate the amplitude and width of the Doppler ridges.  Finally, although we evaluate our pipeline with human motion, the underlying principle is reflector-centric. In other words, our proposed principle can easily extend to other moving reflectors such as vibrating or rotating machinery or other quadrupeds besides human sensing.

We next present the contributions of the paper in more detail.

\textbf{Statement of Contributions:}

1. In this paper, we develop a novel compact and efficient parametrization of WiFi spectrograms as a superposition of a small number of dominant motion ridges, where the Doppler ridges of each major reflector are described by its speed, geometry factor, ridge amplitude, and smear. Moreover, we methodically establish that the problem space can lend itself to a low-dimensional representation through an extensive study of diverse human activities using tools from computer vision. This then provides the tractable foundation for our subsequent ML design.

2. Building on our efficient parametric spectrogram modeling, we propose a physics-informed autoencoder pipeline with an explicitly designed bottleneck to include physically meaningful parameters. Reflector speeds are compactly parametrized using Fourier coefficients, while a handful of parameters determine the geometry factor. A fixed differentiable RF forward model then reconstructs the spectrogram from the bottleneck, recombining the reflector speeds and geometry factor and constraining the bottleneck to stay physically meaningful. 

3. We then introduce a synthetic-to-real training framework to enable the neural network to learn meaningful physical variables that explain a spectrogram, the alternative of which is a non-trivial task that involves collecting multi-dimensional labeled real WiFi data for training purposes. Our proposed synthetic spectrogram generator provides complete labels for reflector speeds, geometry factor, ridge amplitudes, and smears while producing noisy/clean spectrogram pairs. This framework enables the complete model and its hyperparameters to be learned without using real WiFi measurements, while also teaching the model to work with and denoise real WiFi spectrograms.

4. We extensively validate the proposed framework under both known and time-varying geometry using both independently generated synthetic test sets as well as $31$ real WiFi experiments. The real WiFi evaluation includes geometry-varying walking trials from subjects across multiple locations, hand-induced motions, as well as testing on an existing walking dataset of $26$ fixed-angle walking trials from $13$ subjects across two locations. We note that no real WiFi measurements are used for training or hyperparameter selection. The results confirm that the proposed pipeline can robustly estimate speed profile, underlying geometry, Doppler ridge amplitude, and ridge width across all the experiments. For varying-geometry cases, it achieves speed NMSEs of \(0.027\) on the synthetic test set and \(0.166\) on the real WiFi test set, reducing the NMSE of the strongest baseline by \(86.3\%\) and \(41.8\%\), respectively. Moreover, the proposed model demonstrates strong geometry-estimation performance, whereas existing baselines struggle with geometry estimation and rely on assumed values, resulting in substantial estimation errors.

%%%%%%%%%%%%%%%% Sec 2  %%%%%%%%%%%%%%%%%%%
\section{Problem Formulation}
\label{sec:probformulation}

In this paper, we consider the challenging problem of estimating the speeds of the significant moving reflectors using a single WiFi TX-RX link. One key difficulty is that the dominant ridge(s) in a WiFi spectrogram are not direct measurements of physical speed(s). Instead, the observed frequency of each ridge is governed by the product of the reflector speed and an instantaneous physical geometry factor. For instance, a similar Doppler frequency can be produced by fast motion under a weak geometric projection or by slower motion under a stronger geometric projection. Second, practical limitations such as limited bandwidth and noise can smear/obscure the spectrogram, further complicating speed/geometry extraction.

In this section, we start by summarizing the physical relationship between reflector motion and Doppler frequency, followed by posing the problem of interest to this paper.

\textbf{Ridge Definition:} In this paper, we define a ridge as a high-energy Doppler trace in the spectrogram.

\subsection{Doppler Signatures in WiFi Spectrograms -- A Primer}
\label{subsec:motion_to_doppler}

Let a WiFi transmitter and receiver be located at $\mathbf{p}_{\mathrm{tx}},\mathbf{p}_{\mathrm{rx}}\in\mathbb{R}^3$, respectively, and let $\mathbf{p}_m(t)$ denote the position of the $m^{\mathrm{th}}$ moving reflector (e.g., a body part) at time $t$. Let $M$ denote the number of moving reflectors in the scene. As established in the literature \cite{korany2019xmodal, torun2025fast}, the narrowband received complex baseband signal can be modeled as the sum of a static component and the reflections from moving reflectors:
\begin{equation}
\label{eq:complex_baseband}
    y(t)
    =
    y_{\mathrm{s}}
    +
    \sum_{m=1}^{M}
    \alpha_m(t)
    \exp\left(
        -j\frac{2\pi}{\lambda}d_m(t)
        +j\varphi_m
    \right),
\end{equation}
where $y_{\mathrm{s}}$ captures the contribution of the direct path and reflections from static objects, $\alpha_m(t)$ is the effective amplitude of the signal reflected from reflector $m$, $\lambda$ is the wavelength, and $\varphi_m$ is a constant phase offset determined by the initial phase of the corresponding path. The propagation path length associated with reflector $m$ is
\begin{equation}
\label{eq:reflector_range}
    d_m(t)
    =
    \left\|
        \mathbf{p}_m(t)-\mathbf{p}_{\mathrm{tx}}
    \right\|_2
    +
    \left\|
        \mathbf{p}_m(t)-\mathbf{p}_{\mathrm{rx}}
    \right\|_2.
\end{equation}
Motion of reflector $m$ causes $d_m(t)$ to vary and hence changes the phase of the corresponding reflected component in \eqref{eq:complex_baseband}. Because Doppler frequency is determined by the rate of this phase change, we first compute the rate of change of the reflected path length. Define the unit vectors pointing from the transmitter and receiver toward reflector $m$ as
\begin{equation}
\label{eq:unit_vectors}
    \mathbf{u}_{\mathrm{tx},m}(t)
    =
    \frac{
        \mathbf{p}_m(t)-\mathbf{p}_{\mathrm{tx}}
    }{
        \left\|
            \mathbf{p}_m(t)-\mathbf{p}_{\mathrm{tx}}
        \right\|_2
    },
    \qquad
    \mathbf{u}_{\mathrm{rx},m}(t)
    =
    \frac{
        \mathbf{p}_m(t)-\mathbf{p}_{\mathrm{rx}}
    }{
        \left\|
            \mathbf{p}_m(t)-\mathbf{p}_{\mathrm{rx}}
        \right\|_2
    }.
\end{equation}
Differentiating Eq.~\eqref{eq:reflector_range} then gives the rate of change of the reflected path length:
\begin{align}
\label{eq:path_length_derivative}
    \dot{d}_m(t)
    &=
    \frac{d}{dt}d_m(t)
    \nonumber\\
    &=
    \left(
        \mathbf{u}_{\mathrm{tx},m}(t)
        +
        \mathbf{u}_{\mathrm{rx},m}(t)
    \right)^{\top}
    \dot{\mathbf{p}}_m(t).
\end{align}
The phase of the $m^{\mathrm{th}}$ reflected component in \eqref{eq:complex_baseband} is $-2\pi d_m(t)/\lambda+\varphi_m$. Its signed instantaneous Doppler frequency is therefore
\begin{align}
\label{eq:signed_doppler}
    \widetilde{f}_{D,m}(t)
    &=
    \frac{d}{dt}
    \left(
        -\frac{2\pi}{\lambda}d_m(t)
        +
        \varphi_m
    \right)
    \nonumber\\
    &=
    -\frac{2\pi}{\lambda}\dot{d}_m(t)
    \nonumber\\
    &=
    -\frac{2\pi}{\lambda}
    \left(
        \mathbf{u}_{\mathrm{tx},m}(t)
        +
        \mathbf{u}_{\mathrm{rx},m}(t)
    \right)^{\top}
    \dot{\mathbf{p}}_m(t) (\mathrm{rad/s})\nonumber\\
    &=
    -\frac{1}{\lambda}
    \left(
        \mathbf{u}_{\mathrm{tx},m}(t)
        +
        \mathbf{u}_{\mathrm{rx},m}(t)
    \right)^{\top}
    \dot{\mathbf{p}}_m(t) (\mathrm{Hz}).
\end{align}

Let
\begin{equation}
\label{eq:speed_direction}
    v_m(t)
    =
    \left\|
        \dot{\mathbf{p}}_m(t)
    \right\|_2
\end{equation}
denote the speed of reflector $m$. Whenever $v_m(t)\ne 0$, define its instantaneous direction of motion as
\begin{equation}
\label{eq:heading_direction}
    \mathbf{h}_m(t)
    =
    \frac{
        \dot{\mathbf{p}}_m(t)
    }{
        v_m(t)
    },
    \qquad
    \left\|
        \mathbf{h}_m(t)
    \right\|_2
    =
    1.
\end{equation}
Thus,
$\dot{\mathbf{p}}_m(t)=v_m(t)\mathbf{h}_m(t)$.

In most WiFi sensing work, the spectrograms are constructed from real-valued WiFi motion signals derived from channel state information (CSI) power or receiver phase difference \cite{wang2015understanding, wang2017phasebeat}. Their two-sided spectra are conjugate symmetric, hence only the nonnegative-frequency portion is considered. Accordingly, the Doppler sign is not preserved; hence, we define the observed Doppler magnitude due to a moving reflector $m$ as
\begin{equation}
\label{eq:doppler_magnitude}
    f_{D,m}(t)
    \triangleq
    \left|
        \widetilde{f}_{D,m}(t)
    \right|.
\end{equation}
Combining Eqs.~\eqref{eq:signed_doppler}, \eqref{eq:heading_direction}, and \eqref{eq:doppler_magnitude} gives
\begin{equation}
\label{eq:fd_product_general}
    f_{D,m}(t)
    =
    \frac{\psi_m(t)v_m(t)}{\lambda},
\end{equation}
where
\begin{equation}
\label{eq:geometry_factor}
    \psi_m(t)
    =
    \left|
        \left(
            \mathbf{u}_{\mathrm{tx},m}(t)
            +
            \mathbf{u}_{\mathrm{rx},m}(t)
        \right)^{\top}
        \mathbf{h}_m(t)
    \right|,
    0\leq \psi_m(t)\leq 2,
\end{equation}
is the instantaneous geometry factor. 

\textbf{Remark 1:} Throughout this paper, $\psi_m(t)$ denotes this nonnegative geometry factor.

When a short-time Fourier transform (STFT) is applied along the time dimension of the received WiFi CSI, and when $f_{D,m}(t)$ varies sufficiently slowly within each STFT window, the energy associated with body component $m$ is concentrated near
\begin{equation}
\label{eq:stft_ridge_frequency}
    f
    =
    f_{D,m}(t)
    =
    \frac{
        \psi_m(t)v_m(t)
    }{
        \lambda
    }.
\end{equation}

Eq.~\eqref{eq:stft_ridge_frequency} shows the underlying physical relationship governing the center frequencies of Doppler-induced spectrogram ridges.

\subsection{The Product Ambiguity}
\label{subsec:product_ambiguity}

For a single isolated ridge due to a reflector in the vicinity of a WiFi link, define the measured ridge observation as
\begin{equation}
\label{eq:ridge_freq}
    r_m(t) \triangleq \lambda f_{D,m}(t)=\psi_m(t)v_m(t).
\end{equation}
Infinitely many speed--geometry pairs can explain the same observation:
\begin{equation}
\label{eq:ambiguity_family}
    \psi_x(t)=\frac{\psi_m(t)}{c(t)},\quad
    v_x(t)=c(t)v_m(t),
\end{equation}
for any positive function $c(t)$ such that $0\leq\psi_x(t)\leq 2$. Therefore, purely theoretical bandwidth or ridge-based estimators cannot recover true speed without either assuming $\psi(t)$, measuring geometry externally, using multiple independent links, or imposing strong assumptions on the motion context. This is the fundamental weakness of methods that assume a fixed geometry factor, as subjects in the wild can change position, heading, or path relative to the link. The following lemma summarizes this ambiguity. 

\begin{lemma}
\label{lemma:nonidentifiability}
Consider a noiseless single-component spectrogram with one ridge centered at $f_D(t)=\psi(t)v(t)/\lambda$. Suppose that $\psi(t)>0$ and $v(t)>0$ over an interval $\mathcal{T}$. For any positive smooth function $c(t)$ satisfying $0<\psi(t)/c(t)\leq 2$, the pair $(\psi(t)/c(t),c(t)v(t))$ produces exactly the same Doppler frequency over $\mathcal{T}$. Hence the true speed $v(t)$ is not identifiable from the observation.
\end{lemma}

In addition to this fundamental ambiguity, practical WiFi spectrograms are imperfect: limited bandwidth and sampling effects, as well as similar but nonidentical Doppler contributions from different points on an extended moving reflector, can smear Doppler signatures, further complicating speed/geometry extraction. As a result, for real WiFi spectrograms, the dominant motion ridges occupy a frequency band and we do not observe numerous reflectors' individual speeds as infinitely narrow tracks, a phenomenon which we shall refer to as the ``smearing'' effect in the rest of the paper. Thus, training a generic neural network directly on spectrograms may not suffice to address the problem of interest. This motivates our proposed physics-informed ML pipeline, where speed and geometry are explicitly represented in the network bottleneck and constrained through spectrogram reconstruction using a differentiable RF forward model, as we shall see.

%%%%%%%%%%%%%%%% Sec 3   %%%%%%%%%%%%%%%%%%%
\section{Parametric Forward Modeling for WiFi Spectrograms}
\label{sec:parametric_spectrogram}
To enable the physics-informed learning framework developed in the next section, we first establish an efficient parametric model of WiFi spectrograms. Despite the efficiency of this representation, the resulting inverse problem can still remain intractable if a large number of components were required. On the other hand, the large number of moving points on the human body can intuitively be expected to exhibit substantial redundancy in their motion and, consequently, in their contributions to the resulting spectrogram. This section then methodically validates this hypothesis through an extensive study of diverse human activities using tools from computer vision, establishing the low-dimensional, tractable foundation for our subsequent ML design.

\subsection{Ridge Modeling of WiFi Spectrograms}
\label{subsec:spectrogram_model}
We next introduce a compact parametric model for the dominant ridges of a WiFi spectrogram. Each dominant motion component is represented by a localized frequency-band kernel, whose center frequency is determined by the reflector speed and geometry factor, while its amplitude and width capture the relative reflection strength and Doppler smearing, respectively.

More specifically, let $S(f,t)$ denote a spectrogram obtained by applying STFT to a received WiFi CSI signal, and let $K_{w_m(t)}(\Delta f)$ denote a normalized, symmetric kernel centered at $\Delta f=0$
\begin{equation}
\label{eq:kernel_normalization}
    \max_{\Delta f\in\mathbb{R}}
    K_{w_m(t)}(\Delta f)
    =
    1,
\end{equation}
where $w_m(t)>0$ characterizes the effective frequency width of the corresponding ridge.

A spectrogram can then be modeled as
\begin{equation}
\label{eq:spectrogram_forward}
    S(f,t)
    \approx
    \sum_{m=1}^{M}
    a_m(t)
    K_{w_m(t)}
    \left(
        f-\frac{\psi_m(t)v_m(t)}{\lambda}
    \right)
    +
    \eta(f,t),
\end{equation}
where $a_m(t)\geq 0$ denotes the effective reflection strength of component $m$ or, equivalently, the relative ridge amplitude, and $\eta(f,t)$ is the spectrogram noise due to residual DC leakage, unrelated minor motions in the environment, measurement/hardware noise, multipath interference, and possible signal-processing artifacts. Jointly, $a_m(t)$ and $w_m(t)$ describe the relative amplitude and frequency smearing amount of a ridge, whereas the speed and geometry define the location of the corresponding ridge in the frequency domain. We note that for a basic STFT with sliding rectangular time windows, this kernel would become the well-known sinc or sinc$^2$ function.

\begin{figure*}[t]
    \centering
    \includegraphics[width=\textwidth]{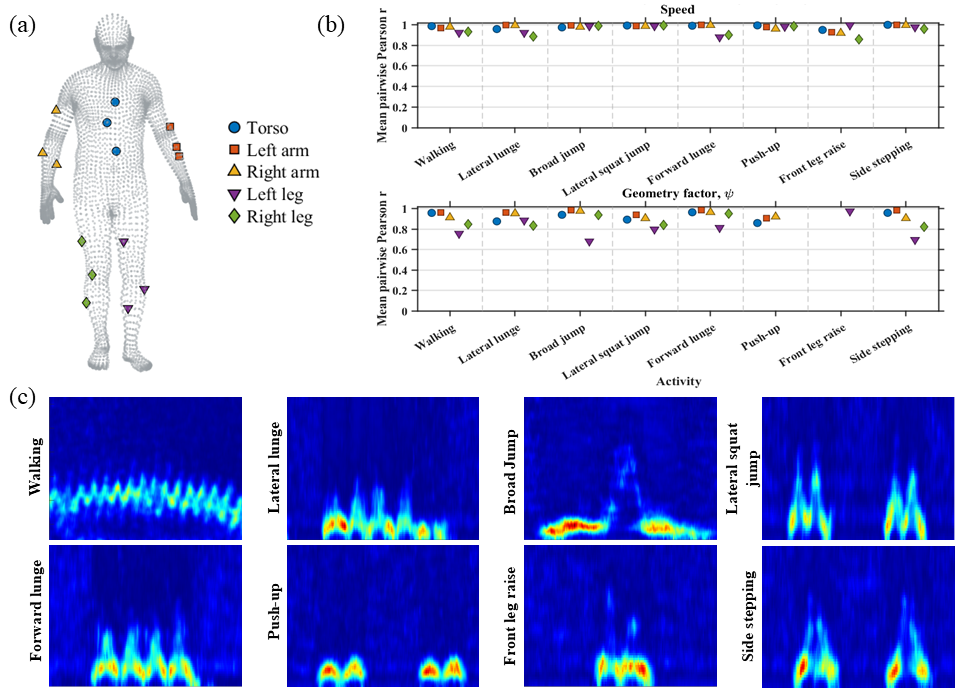} 
    \vspace{8pt}
    \caption{Human mesh recovery (HMR) confirms the low-dimensional motion structure enabling our compact spectrogram model. -- (a) Recovered human mesh and three sample points randomly selected from five main body parts of a human, (b) Mean pairwise Pearson correlation coefficient $r$ among the chosen sample points within each body part for speed and $\psi$, evaluated across eight activities. Missing $\psi$ markers are due to insufficient significant movement of certain body parts for specific activities, (c) Sample real WiFi spectrograms across various activities from an open-source dataset~\cite{cai2020teaching,gym_dataset}, further confirming the low-dimensional structure observed in the HMR analysis, with most of the Doppler energy concentrated around no more than three distinct motion components.}
    \label{fig:fig1}
    \vspace{8pt}
\end{figure*}

Eq.~\eqref{eq:spectrogram_forward} is a parametrized forward model of a WiFi spectrogram rather than an exact definition. The reflected paths in Eq.~\eqref{eq:complex_baseband} are superposed at the signal level before any non-linear operations. Consequently, the exact spectrogram can contain interference and cross-reflection terms that are not expressible as a simple sum of independent ridges. Nonetheless, Eq.~\eqref{eq:spectrogram_forward} provides us with a strong and effective model to explain and recreate an approximation of a spectrogram from physically meaningful parameters. More specifically, $v_m(t)$ and $\psi_m(t)$ jointly determine the center frequency of the $m^{\mathrm{th}}$ ridge, $a_m(t)$ determines the ridge relative amplitude, and $w_m(t)$ determines the effective frequency smear amount. The inverse problem of interest is then to infer the underlying physical speeds $\{v_m(t)\}_{m=1}^{M}$ as well as the geometry factors $\{\psi_m(t)\}_{m=1}^{M}$ from an observed spectrogram $S(f,t)$.

\subsection{Establishing a Tractable Low-Dimensional Representation}
\label{subsec:limitM}

Eq.~\eqref{eq:spectrogram_forward} represents a WiFi spectrogram as a superposition of parametrized motion ridges. The tractability of this representation, however, critically depends on the number of distinct motion components required to explain the spectrogram. Human motion, on the other hand, exhibits substantial structural redundancy: many moving points share highly correlated motion. We methodically establish this low-dimensional structure through an extensive study of diverse human activities using computer vision, providing a tractable foundation for our subsequent ML design.

More specifically, we consider an open-source dataset of videos spanning eight common activities with diverse human motion patterns: walking, lateral lunge, broad jump, lateral squat jump, forward lunge, push-up, front leg raise, and side stepping~\cite{cai2020teaching}. We apply a human mesh recovery (HMR) algorithm \cite{kanazawa2018end} to each video to recover the motion of the human body parts over time. We then randomly sample three points from five main body parts of a human, as illustrated in Fig.~\ref{fig:fig1}(a). Next, we compute the speed of each point and the $\psi$ for a nearby imaginary WiFi TX-RX link. In order to compute the speeds in m/s instead of pixels/frame, we utilize the video frames per second information as well as the known path length at subject depth. We note that, for a robust analysis, $\psi$ is calculated only when the instantaneous point speed is at least $0.25~\mathrm{m/s}$.

For each activity and main body part, Fig.~\ref{fig:fig1}(b) reports the mean of the three unique pairwise Pearson correlation coefficients among the selected points. Across all combinations, the average speed correlation is $0.967$ (median $0.984$; minimum $0.858$), while the average deviation from the body part mean speed is $0.028~\mathrm{m/s}$. As can be seen, the sampled points within each body part show strongly correlated speeds. Similarly, for $\psi$ correlations, a mean of $0.898$ and a median of $0.917$ were observed. Geometry-factor similarity is particularly strong for the torso and arms, but is weaker, especially for the left leg. We note that the videos we used are from the right-side view; hence, the HMR algorithm is less reliable for the less visible left leg due to right leg occupancy of the frames, which explains the degradation.

Fig.~\ref{fig:fig1}(c) further confirms the low dimensionality by showing various real WiFi spectrograms from the aforementioned activities. As can be seen, each WiFi spectrogram can be explained by no more than three visually distinct ridges.

Overall, our analysis establishes that human motion-induced WiFi spectrograms can be effectively represented by a small number of dominant motion components with distinct speeds. Based on these results, we adopt $M\in\{0,1,2,3\}$ as our modeling assumption, yielding the compact, tractable representation underlying our ML design.

\begin{figure*}[t]
    \centering
    \includegraphics[width=\textwidth]{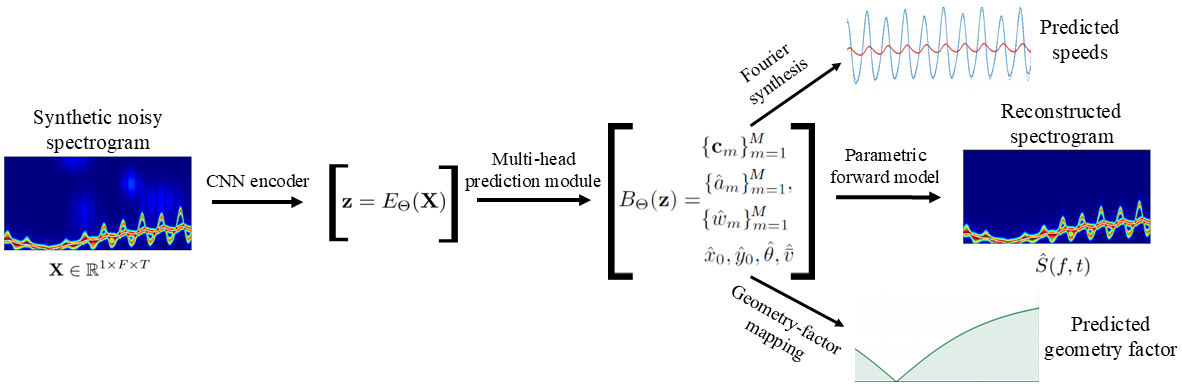}
    \vspace{-8pt}
    \caption{Overview of the proposed physics-informed autoencoder. A CNN maps the input WiFi spectrogram $\mathbf{X}$ to a structured physical bottleneck containing: Fourier coefficients $c_m$ to represent reflector speeds, scenario-related geometry parameters  $\hat x_0,\hat y_0,\hat\theta,\hat{\bar v}$ which determine the geometry factor $\hat\psi(t)$, ridge amplitudes $a_m$, and ridge widths $w_m$. In addition to parameter-level supervision, the reconstruction loss couples all estimated physical parameters through the physics-informed RF forward model, requiring their joint prediction to reproduce the target spectrogram.}
    \label{fig:fig2}
    \vspace{-8pt}
\end{figure*}

%%%%%%%%%%%%%%%% Sec 4   %%%%%%%%%%%%%%%%%%%
\section{Physics-Informed Autoencoder for WiFi Spectrograms}
\label{sec:systemdesign}

In this section, we introduce our physics-informed autoencoder for recovering the speeds and geometry from WiFi spectrograms. The key design is a structured, interpretable bottleneck that explicitly represents the physical parameters of the spectrogram model developed in Sec.~\ref{sec:parametric_spectrogram}, including reflector speeds, geometry, ridge amplitudes, and ridge smearing. Rather than employing a learned decoder, we use a fixed differentiable RF forward model to reconstruct the spectrogram directly from these estimated physical parameters. In addition to parameter-level supervision, the resulting reconstruction loss couples all estimated physical parameters through the physics-informed RF forward model, requiring their joint prediction to reproduce the target spectrogram. Fig.~\ref{fig:fig2} provides an overview of the proposed architecture. We next discuss each component in detail.

\subsection{Encoder Structure and Physical Bottleneck}
\label{subsec:bottleneck}

The encoder takes as input $\mathbf{X}$, a single-channel WiFi spectrogram similar to how an autoencoder receives an image,
\begin{equation}
    \mathbf{X}\in\mathbb{R}^{1\times F\times T}.
\end{equation}
where $F$ and $T$ denote the number of frequency and time bins, respectively. A custom CNN encoder processes $\mathbf{X}$ using four stride-two convolutional blocks, each followed by batch normalization and a ReLU activation. Adaptive pooling and a fully connected layer then produce the latent representation:
\begin{equation}
    \mathbf{z}=E_{\Theta}(\mathbf{X})\in\mathbb{R}^{d_z}.
\end{equation}
Unlike a conventional autoencoder, the bottleneck vector is not passed through the decoder as is. Instead, a shared multilayer perceptron followed by task-specific prediction heads map $\mathbf{z}$ to the physical bottleneck that consists of four main parts: speed, ridge amplitude, ridge width, and $\psi$. We formally define the physical bottleneck as
\begin{equation}
\label{eq:bottleneck_output}
    B_{\Theta}(\mathbf{z})
    =
    \left\{
    \{\mathbf{c}_m\}_{m=1}^{M},
    \{\hat a_m\}_{m=1}^{M},
    \{\hat w_m\}_{m=1}^{M},
    \hat x_0,\hat y_0,\hat\theta,\hat{\bar v}
    \right\},
\end{equation}
where $\mathbf{c}_m\in\mathbb{R}^{1+2K}$ contains the Fourier coefficients that parameterize the speed of reflector $m$, $\hat a_m$ denotes its relative ridge amplitude, and $\hat w_m$
denotes its ridge width. The geometry factor is not predicted independently at every time bin. Instead, the geometry head predicts four parameters that describe a walking path: the initial position $(\hat x_0,\hat y_0)$, heading angle $\hat\theta$, and mean translational speed $\hat{\bar v}$. These parameters generate a time-varying trajectory and, subsequently, the shared geometry factor $\hat\psi(t)$. This low-dimensional construction reduces the output dimension and restricts $\hat\psi(t)$ to a physically structured family rather than allowing an unconstrained value at each time instant. We note that the amplitude of the primary reflector is fixed as $\hat a_1=1$, whereas the relative amplitude $\hat a_m\in[0,1]$ of any additional reflector is predicted by the network.

For physical variables, the corresponding prediction heads map an unconstrained network output to a predefined range:
\begin{equation}
\label{eq:sigmoid_map}
    \hat q=q_{\min}+(q_{\max}-q_{\min})\sigma(r_q),
\end{equation}
where $r_q$ is an unconstrained network output. This mapping is applied to the path parameters defining $\psi$, ridge amplitudes, and widths. Finally, the reflector predicted speeds are bounded after the Fourier reconstruction, which is described next.

Similar to our low-dimensional parameterization of $\psi$, we next introduce a compact representation for the time-varying reflector speeds. Directly predicting the speeds at every time bin would substantially increase the bottleneck dimension and limit scalability. We therefore parameterize each reflector's speed using a truncated Fourier series whose fundamental period equals the observation duration:

\begin{equation}
\label{eq:fourier_raw_speed}
    \rho_m(t)=c_{m,0}+\sum_{k=1}^{K}
    \left(
    c_{m,k}^{\cos}\cos\frac{2\pi kt}{T_{\mathrm{win}}}
    +
    c_{m,k}^{\sin}\sin\frac{2\pi kt}{T_{\mathrm{win}}}
    \right),
\end{equation}
and
\begin{equation}
\label{eq:bounded_speed}
    \hat v_m(t)=v_{\max}\sigma\!\left(\rho_m(t)\right).
\end{equation}
where $\mathbf{c}_m=[c_{m,0},c_{m,1}^{\cos},c_{m,1}^{\sin},\ldots,c_{m,K}^{\cos},c_{m,K}^{\sin}]^{\mathsf T}$.

As a design choice, we use $T_{\mathrm{win}}=6$ s. The maximum speed is then set according to the modeled Doppler band,
\begin{equation}
\label{eq:speed_bound}
    v_{\max}=\frac{f_{\max}\lambda}{\psi_{\max}} \approx 3~\mathrm{m/s},
\end{equation}
where $f_{\max}=100$ Hz, $\lambda \approx 0.06$ m, and $\psi_{\max}=2$. Note that for most common activities captured with WiFi spectrograms, $100$ Hz provides a sufficiently large upper bound for the Doppler frequencies associated with the considered human activities, ensuring that every rendered ridge remains within the modeled frequency range. Overall, the proposed Fourier representation enforces a compact structure while retaining sufficient flexibility. We next discuss the full forward path. 

\subsection{Differentiable RF Forward Model}
\label{subsec:forwardmodel}

A key component of our architecture is the replacement of a conventional learned decoder with a fixed, differentiable RF forward model, i.e., a differentiable implementation of the spectrogram parameterization introduced in Sec.~\ref{sec:parametric_spectrogram}. The forward model contains no trainable parameters. Instead, it maps the predicted physical bottleneck $B_{\Theta}(\mathbf z)$ directly to the spectrogram domain through a sequence of analytical operations. This approach enables the reconstruction loss to jointly constrain the predicted physical parameters by requiring them to reproduce the target spectrogram through the RF forward model.

We next discuss the steps in achieving this forward model. First, the $\psi(t)$ prediction is generated through the walking path model. From the predicted starting point, mean translational speed, and heading angle, the trajectory is calculated
\begin{equation}
\label{eq:path_model}
    \hat{\mathbf{p}}(t)=
    \begin{bmatrix}
        \hat x_0\\
        \hat y_0
    \end{bmatrix}
    +
    \hat{\bar v}t
    \begin{bmatrix}
        \cos\hat\theta\\
        \sin\hat\theta
    \end{bmatrix}.
\end{equation}
Let $\hat{\mathbf h}=[\cos\hat\theta,\sin\hat\theta]^{\mathsf T}$ denote the walking direction and let
\begin{align}
    \hat{\mathbf u}_{\mathrm tx}(t)
    &=
    \frac{\hat{\mathbf p}(t)-\mathbf p_{\mathrm tx}}
    {\|\hat{\mathbf p}(t)-\mathbf p_{\mathrm tx}\|_2+\epsilon},
    \\
    \hat{\mathbf u}_{\mathrm rx}(t)
    &=
    \frac{\hat{\mathbf p}(t)-\mathbf p_{\mathrm rx}}
    {\|\hat{\mathbf p}(t)-\mathbf p_{\mathrm rx}\|_2+\epsilon}.
\end{align}
where $\epsilon>0$ is a small constant used for numerical stability. The geometry factor is then computed as
\begin{equation}
\label{eq:decoder_psi}
    \hat\psi(t)
    =
    \left|
    \left(
        \hat{\mathbf u}_{\mathrm tx}(t)
        +
        \hat{\mathbf u}_{\mathrm rx}(t)
    \right)^{\mathsf T}
    \hat{\mathbf h}
    \right|,
    \qquad
    0\leq\hat\psi(t)\leq2.
\end{equation}
Whereas the estimated speeds $\{\hat v_m(t)\}_{m=1}^{M}$ are reconstructed from the predicted
Fourier coefficients using Eqs.~\eqref{eq:fourier_raw_speed} and \eqref{eq:bounded_speed}.

Hence, the predicted ridge locations can be computed as
\begin{equation}
\label{eq:predicted_doppler}
    \hat f_{D,m}(t)
    =
    \frac{\hat\psi(t)\hat v_m(t)}{\lambda}.
\end{equation}

Finally, the parametrized spectrogram is rendered from the physical bottleneck $B_{\Theta}(\mathbf{z})$
\begin{equation}
\label{eq:decoder_spec}
    \hat S(f,t)=
    \mathcal{N}\!\left[
    \sum_{m=1}^{M}
    \hat a_m
    \operatorname{sinc}^{2}\!\left(
    \frac{f-\hat f_{D,m}(t)}
    {\hat w_m/2+\epsilon}
    \right)
    \right],
\end{equation}
where $\mathcal{N}[\cdot]$ denotes column-wise normalization. Thus, the predicted speeds and geometry factor determine the ridge locations, whereas $\hat a_m$ and $\hat w_m$ control their relative strengths and frequency smear amount. The decoder is intentionally restricted to this dominant-ridge structure so that the predicted physical variables remain responsible for explaining the observation. Because every operation is differentiable, reconstruction gradients can propagate through the ridge renderer, Doppler mapping, geometry model, and speed parameterization and reach the encoder and prediction heads.

\begin{figure*}[t]
    \centering
    \includegraphics[width=\textwidth]{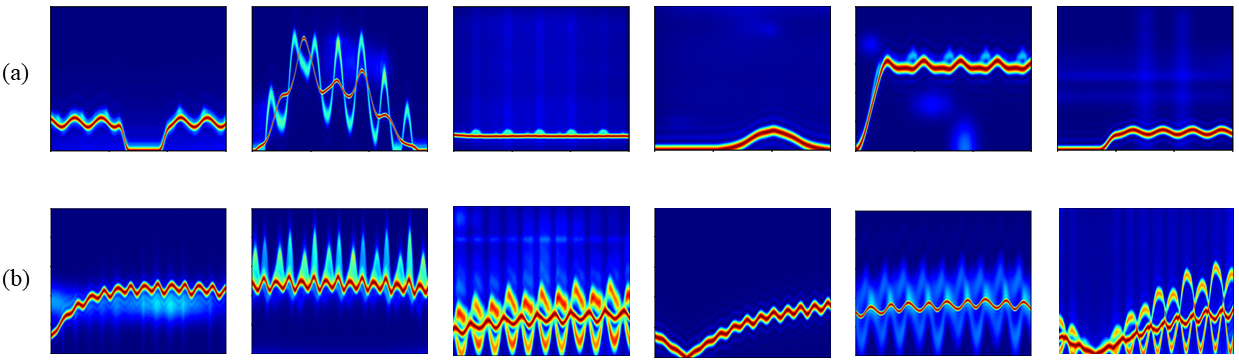}
    \vspace{-8pt}
    \caption{Sampler of the synthetic training dataset. The examples show variation in speeds, geometry factors, relative ridge amplitudes, ridge widths, and structured noise patterns. The corresponding noiseless versions of the spectrograms (not shown) are utilized as reconstruction targets, whereas the noisy spectrograms are provided to the encoder as input. -- (a) Sample spectrograms for the fixed-$\psi$ case, (b) Sample spectrograms for the varying-$\psi$ human walking case.}
    \label{fig:fig3}
    \vspace{-8pt}
\end{figure*}

\subsection{Synthetic Training Data}
\label{subsec:synthetic_training}

Training the proposed bottleneck requires supervision for physical variables that are difficult to obtain concurrently from real WiFi measurements. To achieve such a task, one would require concurrent measurements of reflector speeds and the geometry factor, as well as ground-truth labels for the relative ridge amplitudes and widths, the latter of which is considerably challenging to measure. We therefore construct synthetic spectrograms inspired by the same parametric rendering process employed by our physics-informed decoder. Each generated sample provides the complete physical parameter set $\{\mathbf v,\psi,\mathbf a,\mathbf w\}$ together with an optional noisy input spectrogram and its corresponding clean spectrogram target. We note that by doing so, we allow our proposed model to act like a denoiser for real WiFi spectrograms in addition to being a regressor. We next describe the synthetic spectrogram generator in detail.

For each sample, the generator first randomly draws the component speeds, geometry factor, amplitudes, and ridge widths from predefined ranges and renders a clean spectrogram using the physical model in Sec.~\ref{subsec:forwardmodel}. Structured noise is then added to form the encoder input. As can be seen from the samples provided in Fig.~\ref{fig:fig3}(a), the noise models include uncorrelated noise, smooth two-dimensional clutter, transient vertical column noise, persistent horizontal band noise, low-frequency background noise, localized blob noise, broadband haze noise, and finally mixtures of these. The clean and noisy spectrograms are normalized separately and stored as the reconstruction target and encoder input, respectively. This paired construction trains the model to recover the underlying physical ridge structure rather than reproduce the added noise, which allows for the secondary use case of being a denoiser.

We consider two complementary synthetic trainings as training case studies for the proposed method. The first uses a fixed geometry factor, $\psi(t)=2$, and randomly samples only the reflector speeds and ridge amplitude/widths. Its speed distribution contains numerous speed shapes (discussed in detail in Sec.~\ref{subsec:synthetic_setup}). Fig.~\ref{fig:fig3}(a) showcases samples for the fixed geometry case. By removing geometry variation, this first case isolates the ability of the model to recover component speeds from noisy, smeared, and potentially overlapping ridges.

As a secondary training case, we introduce a context: human walking with various walk geometries. This training case contains nine predefined walking-path scenarios spanning centered and laterally offset horizontal motion, vertical motion, and diagonal motion at $\pm30^\circ$ and $\pm60^\circ$. All scenarios use the same fixed TX-RX pair, while the initial position and heading determine the evolution of $\psi(t)$, i.e., the geometry scenario of interest (more on this in Sec.~\ref{subsec:synthetic_setup}). The sampled speed shapes are inspired by video-pose-extracted body part speeds for human walking. The ridge widths are sampled independently, while the additional reflectors (i.e., legs) exhibit an oscillatory speed around the primary reflector (torso) with approximate anti-phase. We additionally introduce torso-only samples to allow the network to perform well for noisy cases where limbs may not be visible in real WiFi spectrograms. Sample spectrograms for this case can be seen in Fig.~\ref{fig:fig3}(b). We finally note that for both training cases we generate $15{,}000$ training spectrograms and $1{,}500$ separate validation spectrograms. 

\subsection{Loss Structure and Training}
\label{subsec:trainingscheme}
A key benefit of replacing a conventional learned decoder with the fixed, differentiable RF forward model of Sec. IV-B is that it enables the physical bottleneck to be constrained not only through direct parameter-level supervision, but also through the spectrogram reconstruction it produces. Specifically, while individual losses supervise the predicted speed, geometry factor, ridge amplitudes, and ridge widths, the reconstruction loss couples these parameters through the underlying RF physics by further requiring their joint prediction to reproduce the target spectrogram, as we shall next discuss in detail.

\textbf{Reconstruction Loss:}
The reconstruction loss provides the physics-informed component of our training objective by jointly constraining all predicted bottleneck parameters through the RF forward model. Specifically, the predicted parameters are first used to reconstruct the clean spectrogram according to Eq.~\eqref{eq:decoder_spec}, and the resulting reconstruction is compared with the target spectrogram using mean-squared error,

\begin{equation}
\label{eq:ridge_weighted_mse}
    \mathcal{L}_{\mathrm{rec}}
    =
    \frac{1}{BFT}
    \sum_{b,f,t}
    \left(
        \hat S_b(f,t)-S_b(f,t)
    \right)^2,
\end{equation}
where $B$ is the batch size. Predictions are penalized for how much they deviate from the noiseless version of the input spectrogram. Note that only the noisy versions are fed into the network during training.

\textbf{Speed Loss:}
The first additional loss we introduce penalizes predicted speed mismatch. Instead of applying direct MSE loss, we utilize a weighted MSE where weights are the reflector ground truth relative ridge amplitudes, which is directly supervised from the training data. We formally define the speed loss as
\begin{equation}
\label{eq:speed_loss}
    \mathcal{L}_{v}
    =
    \frac{
        \displaystyle
        \sum_{b,m}
        a_{b,m}
        \frac{1}{T}
        \sum_t
        \left(
            \hat v_{b,m}(t)-v_{b,m}(t)
        \right)^2
    }{
        \displaystyle
        \sum_{b,m}a_{b,m}+\epsilon
    }.
\end{equation}
Consequently, an absent or weak ridge does not impose the same speed penalty as compared to what the main reflector or a strong ridge would have.

\textbf{$\psi$ Loss:}
We additionally add direct supervision to the network-predicted $\psi$ with
\begin{equation}
\label{eq:psi_loss}
    \mathcal{L}_{\psi}
    =
    \frac{1}{BT}
    \sum_{b,t}
    \left(
        \hat\psi_b(t)-\psi_b(t)
    \right)^2.
\end{equation}
The $\psi$ loss allows the network to learn how to correctly utilize the $\psi$-corresponding parameters of the bottleneck to draw the correct geometry for an input spectrogram.

\textbf{Amplitude Loss:}
The relative ridge amplitudes are also supervised as
\begin{equation}
\label{eq:amplitude_loss}
    \mathcal{L}_{a}
    =
    \frac{1}{2B}
    \sum_{b,m}
    \left(
        \hat a_{b,m}-a_{b,m}
    \right)^2.
\end{equation}
The main reflector ridge amplitude is fixed at $a_{b,1}=\hat a_{b,1}=1$, hence the network predicts the relative amplitude of the additional components. Note that this does not prevent the reconstruction of an empty spectrogram, since a zero predicted speed effectively removes the corresponding motion ridges.  

\textbf{Smear Loss:}
The ridge widths are supervised using
\begin{equation}
\label{eq:width_loss}
    \mathcal{L}_{w}
    =
    \frac{
        \displaystyle
        \sum_{b,m}
        a_{b,m}
        \left(
            \frac{
                \hat w_{b,m}-w_{b,m}
            }{
                s_w
            }
        \right)^2
    }{
        \displaystyle
        \sum_{b,m}a_{b,m}+\epsilon
    },
\end{equation}
where $s_w$ is chosen as $10$ Hz as a design parameter, which puts the smear loss on a comparable scale. As in the speed loss, amplitude weighting makes the loss aware of when to ignore/put emphasis on the loss depending on the ground-truth reflection amplitude.

\textbf{Coincidence Loss:}
The final objective penalizes persistent coincidence between predicted speed components. Without such a constraint, the network can exploit the reconstruction objective by allowing multiple predicted speed components to collapse to the same or nearly identical speeds, particularly when the secondary reflector amplitude is weak. We therefore introduce a penalty that encourages distinct predicted speed components to remain sufficiently separated. Note that unlike the speed and smear losses, this loss is not amplitude-aware.

Let $\mathcal{P}$ denote the set of unique component pairs and $\mathcal{W}$ the set of overlapping temporal windows. For each component pair $(m,n)$, we first define the mean predicted speed separation within temporal window $\mathcal{I}$ as
\begin{equation}
\label{eq:window_speed_separation}
    \bar d_{b,m,n,\mathcal{I}}
    =
    \frac{1}{\lvert\mathcal{I}\rvert}
    \sum_{t\in\mathcal{I}}
    \left|
        \hat v_{b,m}(t)-\hat v_{b,n}(t)
    \right|.
\end{equation}
The coincidence loss is then defined as
\begin{equation}
\label{eq:coinc_loss}
    \mathcal{L}_{\mathrm{coinc}}
    =
    \frac{
        \displaystyle
        \sum_{b=1}^{B}
        \sum_{(m,n)\in\mathcal{P}}
        \sum_{\mathcal{I}\in\mathcal{W}}
        \left[
            \max\left(
                0,\,
                \delta_v-\bar d_{b,m,n,\mathcal{I}}
            \right)
        \right]^2
    }{
        B\lvert\mathcal{P}\rvert\lvert\mathcal{W}\rvert
    },
\end{equation}
where $\delta_v$ is the desired minimum mean speed separation. We use one-second windows with $50\%$ overlap. Averaging the separation within each window allows brief crossings between the predicted speeds, which can naturally occur, while penalizing long-term collapse of speeds.

The complete training objective is then
\begin{equation}
\label{eq:total_loss}
    \mathcal{L}
    =
    \lambda_{\mathrm{rec}}\mathcal{L}_{\mathrm{rec}}
    +
    \lambda_v\mathcal{L}_{v}
    +
    \lambda_{\psi}\mathcal{L}_{\psi}
    +
    \lambda_a\mathcal{L}_{a}
    +
    \lambda_w\mathcal{L}_{w}
    +
    \lambda_{\mathrm{coinc}}\mathcal{L}_{\mathrm{coinc}},
\end{equation}
where the coefficients balance the importance of each aforementioned loss term. Algorithm~\ref{alg:training_step} additionally summarizes the training steps.

\begin{algorithm}[t]
\caption{Physics-informed training step}
\label{alg:training_step}
\begin{algorithmic}[1]
\REQUIRE Noisy input spectrograms $\mathbf{X}$; clean reconstruction targets $\mathbf{S}$; physical variables $\mathbf{v}$, $\psi$, $\mathbf{a}$, and $\mathbf{w}$
\STATE Encode $\mathbf{z}=E_{\Theta}(\mathbf{X})$
\STATE Predict the physical bottleneck using Eq.~\eqref{eq:bottleneck_output}
\STATE Construct component speeds using Eqs.~\eqref{eq:fourier_raw_speed}--\eqref{eq:bounded_speed}
\STATE Compute the shared geometry factor using Eq.~\eqref{eq:decoder_psi}
\STATE Render $\hat{\mathbf{S}}$ using Eq.~\eqref{eq:decoder_spec}
\STATE Compute $\mathcal{L}$ using Eq.~\eqref{eq:total_loss}
\STATE Update the encoder and bottleneck parameters by backpropagating through the RF forward model
\end{algorithmic}
\end{algorithm}

%%%%%%%%%%%%%%%%% Sec 5 %%%%%%%%%%%%%%%%%%%
\section{Experimental Setup}
\label{sec:experimental_setup}

We design our experiments to answer three main questions: 
\begin{enumerate}
    \item With the geometry factor $\psi$ fixed to isolate the speed-recovery problem as a starting point, how well can the proposed pipeline recover and summarize multiple arbitrary reflector speed trajectories (of any shape) under noisy, smeared, and partially overlapping ridges?  
    \item For the time-varying $\psi$ case involving a human walk, how well can the network jointly predict speed and $\psi$?
    \item How well does our model, which is trained entirely on synthetic spectrograms, perform with real WiFi measurements collected across different motion types, subjects, and sensing conditions?
\end{enumerate}
We next discuss the implementation and experiment details.

\subsection{Model Configurations and Hyperparameters}
\label{subsec:implementation_details}

We restrict each input to a six-second single-channel spectrogram. The frequency range is set to $0$--$100$ Hz with a frequency spacing of $0.1$ Hz, resulting in $F=1001$ frequency bins. The time axis spacing is $0.01$ s, resulting in $T=600$ time bins. We use $\lambda=0.06$ m in both the synthetic generator and the RF forward model to simulate WiFi signals. Input and target spectrograms are independently normalized along the frequency axis at each time instant (column-wise normalization). These create a standard for expected WiFi spectrogram values. Therefore, the same frequency grid, time grid, and normalization are used when preparing the real WiFi spectrograms for testing purposes.

As discussed, we consider two variants. The fixed-geometry model, denoted by \textsc{Fixed-$\psi$}, is trained with $\psi(t)=2$ and models up to $M=2$ reflector-speed components using Fourier order $K=20$. The \textsc{Fixed-$\psi$} case is designed to showcase the denoising, information extraction, and smear-fitting capabilities of the proposed model, as well as its ability to recover speeds of any shape (not just human walking). The walking-context model denoted by \textsc{Varying-$\psi$}, on the other hand, is trained with the nine walking-path scenarios described in Sec.~\ref{subsec:synthetic_training} and is designed to jointly estimate $\psi$ with up to $M=3$ speed components using $K=30$. The \textsc{Varying-$\psi$} case is designed to show the capabilities of the proposed model for untangling speed and $\psi$.

\begin{table}[t]
\centering
\caption{Model and optimization configurations.}
\label{tab:training_configuration}
\footnotesize
\setlength{\tabcolsep}{5pt}
\begin{tabular}{p{0.20\textwidth}p{0.10\textwidth}p{0.15\textwidth}}
\toprule
\textbf{Parameter}
& \textbf{\textsc{Fixed-$\psi$}}
& \textbf{\textsc{Varying-$\psi$}} \\
\midrule
Geometry target & $\psi(t)=2$ & Time-varying $\psi(t)$ \\
Maximum components $M$ & 2 & 3 \\
Fourier order $K$ & 20 & 30 \\
Encoder base channels & 32 & 32 \\
Latent dimension $d_z$ & 768 & 768 \\
Bottleneck hidden dim. & 512 & 512 \\
Dropout & 0.3 & 0.3 \\
Batch size & 32 & 32 \\
Optimizer & AdamW & AdamW \\
Learning rate & $5\times10^{-6}$ & $1\times10^{-4}$ \\
Weight decay & $1\times10^{-6}$ & $1\times10^{-6}$ \\
Epochs & 500 & 372 \\
$(\lambda_{\mathrm{rec}},\lambda_v,\lambda_{\psi})$
& $(1.0,0.5,0.1)$ & $(2.0,0.5,0.5)$ \\
$(\lambda_a,\lambda_w,\lambda_{\mathrm{coinc}})$
& $(0.2,0.2,0.1)$ & $(0.3,0.2,0.0)$ \\
\bottomrule
\end{tabular}
\end{table}

The hyperparameters used to train our model are summarized in Table~\ref{tab:training_configuration}. The model parameters are optimized using AdamW \cite{loshchilov2017fixing} with mixed-precision training. The final training checkpoint is selected using the validation set speed loss improvement. The hyperparameters, including reconstruction objective, loss weights, geometry-head output ranges, and coincidence margin, are selected using the validation set. Model configurations and loss weights are selected using only the validation sets; neither synthetic nor real WiFi test datasets are used during model parameter selection or optimization. Models are implemented in PyTorch and trained on NVIDIA RTX 6000 Ada, which took $72$ seconds per epoch, on average.

\subsection{Synthetic Training and Validation Sets}
\label{subsec:synthetic_setup}

For each variant, training and validation samples are generated independently. The noisy spectrogram is provided to the encoder, whereas the corresponding noise-free spectrogram is used as the reconstruction target. The physical labels used for supervised training are:
\begin{enumerate}
    \item speeds $v_m(t)$
    \item geometry factor $\psi(t)$
    \item ridge relative amplitudes $a_m(t)$
    \item ridge widths $w_m(t)$
\end{enumerate}

For \textsc{Fixed-$\psi$}, the generator draws samples from 10 speed shape families: constant/slowly varying motion, periodic motion, enveloped periodic motion, stop-go pattern, two ramp patterns, single/repeated bursts, random walk, and lastly purely random. The target speed scale is drawn from the ranges $0.05$--$0.70$, $0.40$--$1.20$, $0.90$--$1.80$, and $1.50$--$2.60$ m/s with probabilities $0.45$, $0.30$, $0.15$, and $0.10$, respectively. Component configurations are sampled as a single visible component, a dominant primary component, a walking-like pair, or a general coupled pair with probabilities $0.20$, $0.20$, $0.35$, and $0.25$, respectively. An additional $2\%$ of the samples contain no motion.

For \textsc{Varying-$\psi$}, all speed components follow continuous walking trajectories. The generator samples five overlapping profiles whose torso mean-speed ranges jointly span $0.45$--$1.65$~m/s and whose per-leg stride frequencies span $0.55$--$1.45$~Hz. The torso speed contains gait-synchronous variation, slow drift, and correlated irregularity. The two gait components share a stride clock, are approximately antiphase, and oscillate around the torso trajectory with correlated but nonidentical spans. A sample contains either the torso alone with probability $0.25$ or the torso and both limb components with probability $0.75$.

All nine geometry scenarios use $\mathbf p_{\mathrm{tx}}=[0,-0.25]^{\mathsf T}$ m and $\mathbf p_{\mathrm{rx}}=[0,0.25]^{\mathsf T}$ m. Table~\ref{tab:walking_geometries} provides further detail on the scenarios used to generate different geometry settings. The scenarios are sampled uniformly during data generation.

\begin{table}[t]
\centering
\caption{Walking geometry scenarios used to train \textsc{varying-$\psi$}.}
\label{tab:walking_geometries}
\begin{tabular}{lcc}
\toprule
Scenario & Initial position $(x_0,y_0)$ (m) & Heading \\
\midrule
Centered horizontal & $(1.50,0.00)$ & $0^\circ$ \\
Upper horizontal & $(1.50,1.60)$ & $0^\circ$ \\
Lower horizontal & $(1.50,-1.60)$ & $0^\circ$ \\
Upper vertical & $(3.00,0.55)$ & $90^\circ$ \\
Lower vertical & $(3.00,-0.55)$ & $-90^\circ$ \\
Upward diagonal & $(1.60,-1.80)$ & $30^\circ$ \\
Upward diagonal & $(2.00,-2.60)$ & $60^\circ$ \\
Downward diagonal & $(1.60,1.80)$ & $-30^\circ$ \\
Downward diagonal & $(2.00,2.60)$ & $-60^\circ$ \\
\bottomrule
\end{tabular}
\end{table}

For both cases, the primary-component smear amount is sampled randomly from $\{5,7.5,10,12.5,15,20,30\}$ Hz and the additional-reflector widths from $\{7.5,10,12.5,15,20,25,30,40,50\}$ Hz, with a decreasing probability to the extremities. Finally, the noise model for each spectrogram is sampled from no noise, white Gaussian, two-dimensional clutter, vertical and horizontal band, low-frequency, localized, broadband hazy, and finally a mixture of any two. For each spectrogram an independent noise strength is sampled from $0.05$--$0.10$, $0.10$--$0.20$, or $0.20$--$0.50$ with probabilities $0.50$, $0.35$, and $0.15$, respectively. Note that the spectrograms are column-normalized; hence, $0-1$ is the meaningful amplitude range choice. For training and validation of each case, $15{,}000$ and $1{,}500$ samples are generated, respectively.

\subsection{Testing with Real WiFi}
\label{subsec:real_wifi_data}

We evaluate our proposed pipeline utilizing three different real WiFi experiment types: controlled hand motion experiments with an attached accelerometer for speed ground truth, a human subject walking with concurrent video-derived speed labels under $\psi=2$ geometry, as well as varying $\psi$ geometries. Table~\ref{tab:real_datasets} further shows the real WiFi data summary used for testing. The participants were given a written form with experiment details all of which were conducted according to our IRB committee guidelines. We emphasize that all real WiFi test data are \textbf{excluded} from model training and hyperparameter selection.

\begin{table}[t]
\centering
\caption{Summary of the real-WiFi evaluation sets.}
\label{tab:real_datasets}
\footnotesize
\setlength{\tabcolsep}{3.5pt}
\renewcommand{\arraystretch}{1.15}

\begin{tabular}{@{}lcccc@{}}
\toprule
\textbf{Dataset}
& \textbf{Subjects}
& \makecell{\textbf{No. of}\\\textbf{experiments}}
& \makecell{\textbf{No. of}\\\textbf{locations}}
& \textbf{Ground truth} \\
\midrule
Hand motion
& 1 & 1 & 1 & Accelerometer \\
Fixed-$\psi{=}2$ walk
& 13 & 26 & 2 & Video pose \\
Angle-varying walk
& 2 & 4 & 2 & Video pose \\
\bottomrule
\end{tabular}
\end{table}

Recordings longer than the model duration are evaluated using sliding $6$-s windows with $50\%$ overlap. Predictions from overlapping windows are combined using the average. To reduce boundary effects from the Fourier representation, the first and last $0.25$ sec of each window are excluded from analysis. For the WiFi data collection procedure and preprocessing details, we refer readers to~\cite{korany2019xmodal}.

\subsection{Baselines for Performance Comparison}
\label{subsec:baselines}

To the best of our knowledge, no existing method jointly estimates multiple reflector speeds and $\psi(t)$ from a WiFi spectrogram. Moreover, there is no established method for directly extracting multiple reflector speeds from the spectrogram. In the absence of an existing baseline, we then design three strong non-learning-based ridge-extraction approaches and carefully tailor them to provide competitive baselines for our setting. \textbf{We emphasize that the design of these baselines is itself an additional contribution of this work.} 

\begin{enumerate}
    \item Peak-based extraction: For the $M=2$ experiments, \emph{adaptive peaks} extracts sufficiently dominant and separated peaks from each time column, determines whether a second component is persistently visible throughout an experiment, and associates the detected peaks over time. For the $M=3$ experiments, \emph{separated peaks} detects up to three dominant, well-separated peaks per column and associates them into continuous speeds.
    \item Power-weighted frequency clustering: For $M=2$, \emph{adaptive two-cluster} compares the optimal power-weighted one-cluster and two-cluster representations. A second component is accepted only when the two-cluster solution provides fit improvement. For $M=3$, \emph{weighted spectral $k$-means} computes three power-weighted frequency centroids in each time column and associates the centroids over time, yielding frequency trajectories that are subsequently converted to speeds.
    \item Dynamic programming-based ridge tracking: For $M=2$, \emph{DP two-ridge} uses dynamic programming to find the maximum-score continuous ridge under a frequency-jump penalty. It then locally suppresses candidates around the first ridge and conditionally accepts a sufficiently strong and separated second ridge if detected. For $M=3$, \emph{residual-ridge DP} repeats the dynamic-programming extraction and local-suppression procedure until three continuous ridges are obtained.
\end{enumerate}
We note that these baselines are inspired by past work, albeit not on WiFi spectrogram ridge extraction, that target the directly observable Doppler ridge trajectories. These methods are well-established in classical time-frequency ridge extraction and dynamic path optimization as well as weighted least-squares clustering \cite{iatsenko2016extraction, lloyd1982least}. We adapt each family to our spectrogram-based problem setting and further to the $M=2$ and $M=3$ settings using visibility thresholds and temporal signal associations for limb tracking. These adaptations are specifically designed to give each baseline its strongest variant for the WiFi spectrograms. For the remainder of this manuscript, we correspondingly use the naming convention as: \emph{adaptive peaks}, \emph{adaptive two-cluster}, and \emph{DP two-ridge} for $M=2$; and \emph{separated peaks}, \emph{weighted spectral $k$-means}, and \emph{residual-ridge DP} for $M=3$.

For each baseline $b$, the extracted frequency $\hat f_b(t)$ is converted to speed in m/s according to
\begin{equation}
\label{eq:baseline_speed_conversion}
    \hat v_b(t)
    =
    \frac{\lambda_{\mathrm{WiFi}}}{\psi_{\mathrm{base}}}
    \hat f_b(t),
\end{equation}
where $\lambda_{\mathrm{WiFi}}=0.06$ m is the wavelength and $\psi_{\mathrm{base}}=2$. As there is no existing method to establish a baseline for $\psi$, the baseline assumes the commonly-taken $\psi = 2$, which will be correct for \textsc{Fixed-$\psi$} experiments, but will impact the performance for the general unknown $\psi$ cases.

The estimation performance degradation in the unknown case is therefore especially valuable as it provides a degradation measure for the consequences of treating an unknown geometry factor as $\psi=2$. All baseline methods as well as the proposed pipeline use the same temporal windows, evaluation masks, and sliding window overlap combination procedures for a fair comparison.

\subsection{Evaluation Metrics}
\label{subsec:evaluation_metrics}

Our primary performance metric is Mean Absolute Error (MAE) in m/s, as it directly states the average magnitude of the speed estimation error. Let $\mathcal{T}_r$ denote the evaluated time bins of recording $r$, whose duration is $\geq6$ sec. For a ground-truth speed $v_r(t)$ and estimate $\hat v_r(t)$, the metrics per experiment reported are
\begin{align}
\label{eq:recording_mae}
    \operatorname{MAE}_r
    &=
    \frac{1}{\lvert\mathcal{T}_r\rvert}
    \sum_{t\in\mathcal{T}_r}
    \left|\hat v_r(t)-v_r(t)\right|,
    \\
\label{eq:recording_rmse}
    \operatorname{RMSE}_r
    &=
    \sqrt{
    \frac{1}{\lvert\mathcal{T}_r\rvert}
    \sum_{t\in\mathcal{T}_r}
    \left(\hat v_r(t)-v_r(t)\right)^2
    },
    \\
\label{eq:recording_nmse}
    \operatorname{NMSE}_r
    &=
    \frac{
    \sum_{t\in\mathcal{T}_r}
    \left(\hat v_r(t)-v_r(t)\right)^2
    }{
    \sum_{t\in\mathcal{T}_r}
    v_r(t)^2
    }.
\end{align}
In addition to MAE, we report root mean squared error (RMSE) in m/s, and the normalized mean squared error (NMSE), as secondary metrics. These additional metrics penalize larger errors more heavily than MAE. NMSE in particular expresses squared error relative to the ground-truth speed.

For a dataset with $R$ experiments, we first compute each metric separately for every experiment and then report the mean across all experiments. Predicted speeds that have no intrinsic ordering (such as legs of a human) are matched only for evaluation with a permutation search (direct match or swapped version). Framewise reassignment is not permitted, and the permutation is performed once per experiment in the same way for proposed ML predictions as well as baseline predictions.

%%%%%%%%%%%%%%%% Sec 6  %%%%%%%%%%%%%%%%%%%
\section{Experimental Results}
\label{sec:results}
In this section, we extensively evaluate the performance and capabilities of our proposed pipeline. We begin with fixed-$\psi$ scenarios to isolate and assess speed-estimation performance. This setting also provides the baselines with their most favorable operating conditions. We then move to the more challenging varying-$\psi$ scenarios, where the speeds and $\psi$ must be jointly estimated.

\subsection{Fixed-$\psi$ -- Parameter Estimation on Synthetic Data}
\label{subsec:results_fixed_psi_synthetic}

We begin by evaluating the performance on a synthetic dataset, which is particularly useful for three reasons. First, the synthetic dataset allows us to consider speeds with a wide variety of forms and shapes. Second, it allows us to showcase the pipeline’s performance in estimating the other ridge parameters, i.e., the corresponding amplitudes and widths. Finally, it provides readily available ground truth for direct performance evaluation.

More specifically, we evaluate the performance on an independently generated synthetic test set of $500$ spectrograms, as summarized in Table~\ref{tab:fixed_psi_synthetic_results}. As can be seen, our proposed Fixed-$\psi$ model achieves main and secondary reflector MAEs of 0.034 and 0.085 m/s, respectively. In comparison, the strongest baseline, DP two-ridge, yields MAEs of 0.051 and 0.249 m/s, corresponding to approximately 50\% and 193\% higher errors, respectively. Table~\ref{tab:fixed_psi_synthetic_results} further reports the RMSE and NMSE results, which are consistent with the MAE trends. Importantly, the model also demonstrates its ability to recover the other ridge parameters, as summarized in Table~\ref{tab:fixed_psi_synthetic_ridge_params}, achieving an amplitude MAE of 0.085 and main and secondary smear-width NMSEs of 0.008 and 0.068, respectively. This additional estimation capability is not available with the baseline methods.
 
Finally, the proposed pipeline also identifies the exact number of visible reflectors in 82.6\% of all 500 samples, compared with 72.0\% for DP two-ridge, 71.0\% for adaptive two-cluster, and 56.0\% for adaptive peaks.

\begin{table}[t]
\centering
\caption{Fixed-$\psi$ speed-estimation performance on the synthetic test set. Note that the baselines are specifically designed to provide a strong comparison under this favorable Fixed-$\psi$ setting. MAE and RMSE are in m/s.}
\label{tab:fixed_psi_synthetic_results}
\footnotesize
\setlength{\tabcolsep}{4.5pt}
\begin{tabular}{lcccccc}
\toprule
& \multicolumn{3}{c}{\textbf{Main reflector}}
& \multicolumn{3}{c}{\textbf{Secondary reflector}} \\
\cmidrule(lr){2-4}\cmidrule(lr){5-7}
\textbf{Method}
& \textbf{MAE} & \textbf{RMSE} & \textbf{NMSE}
& \textbf{MAE} & \textbf{RMSE} & \textbf{NMSE} \\
\midrule
DP two-ridge
& 0.051 & 0.208 & 0.051
& 0.249 & 0.534 & 0.272 \\
Adaptive peaks
& 0.187 & 0.448 & 0.237
& 0.306 & 0.581 & 0.323 \\
Adaptive two-cluster
& 0.301 & 0.547 & 0.354
& 0.922 & 1.205 & 1.388 \\
\textbf{Proposed FIXED-$\psi$}
& \textbf{0.034} & \textbf{0.058} & \textbf{0.004}
& \textbf{0.085} & \textbf{0.125} & \textbf{0.015} \\
\bottomrule
\end{tabular}
\end{table}

\begin{table}[t]
\centering
\caption{Fixed-$\psi$ ridge amplitude and smear-width estimation performance on the synthetic test set. Our proposed method can robustly estimate these parameters, while no corresponding baseline estimation method is available.}
\label{tab:fixed_psi_synthetic_ridge_params}
\footnotesize
\setlength{\tabcolsep}{5pt}
\begin{tabular}{lc}
\toprule
\textbf{Estimated parameter} &  \textbf{NMSE} \\
\midrule
Ridge relative amplitude
 & 0.064 \\
Main ridge smear width
 & 0.008 \\
Secondary ridge smear width
 & 0.068 \\
\bottomrule
\end{tabular}
\end{table}

\subsection{Fixed-$\psi$ Speed Estimation with Real WiFi}
\label{subsec:results_fixed_psi_real}

We next test our trained \textsc{Fixed-$\psi$} model directly on a specifically designed real WiFi experiment, beginning our evaluation of synthetic-to-real data generalization. During the experiment, an accelerometer is attached to a moving hand, which provides the speed ground truth. Because the hand is the only intentionally moving reflector, we use the model’s first predicted speed component for evaluation and apply the same criterion to all baselines.

\begin{figure}[t]
\centering
\includegraphics[width=.5\textwidth]{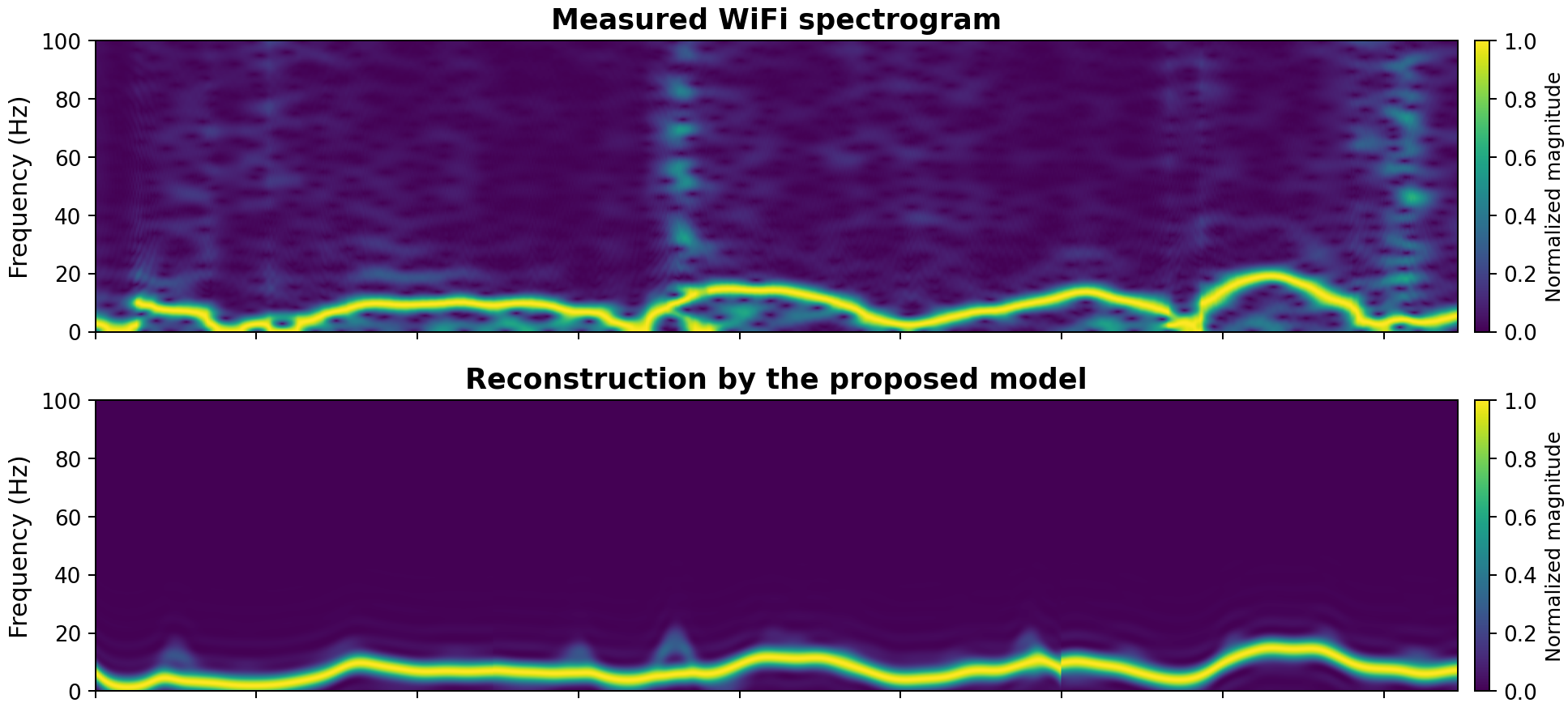}
\caption{Fixed-$\psi$ speed estimation in the controlled real-WiFi hand-motion experiment: the measured WiFi spectrogram and its reconstruction by the proposed model. The strongest baseline has a 26.2\% higher speed MAE than our proposed model.}
\label{fig:fixed_psi_real_hand}
\end{figure}

During the experiment, a stationary subject moved their hand with the palm facing towards the link, while the rest of the body remained still. Our proposed model obtained an MAE of 0.0715 m/s, while the strongest baseline, DP two-ridge, had a 26.2\% higher MAE. Fig.~\ref{fig:fixed_psi_real_hand} additionally showcases the results for this experiment. As shown, the proposed pipeline results in a denoised representation of the input spectrogram that preserves only the Doppler-induced ridge.

\subsection{Varying-$\psi$ Speed and $\psi$ Estimation on Synthetic Data}
\label{subsec:results_varying_psi_synthetic}
The previous two subsections focused on the case of fixed-$\psi$. This section and the next one then extensively investigate the  varying-$\psi$ cases, enabling us to evaluate the joint estimation of both speed and $\psi$.  We then start by considering the joint estimation performance on a synthetic dataset in this section. This, in particular, allows us to also assess the performance of the proposed pipeline when estimating the amplitude and width of the ridges as well. 

More specifically, we start by evaluating on 500 independently generated walking spectrograms spanning the nine geometries in Table~\ref{tab:walking_geometries}. Table~\ref{tab:varying_psi_synthetic_results} summarizes the results for this section. Our proposed model achieves an all-visible-speed MAE of 0.104~m/s, compared with 0.319~m/s for residual-ridge DP, 0.352~m/s for separated peaks, and 0.525~m/s for weighted spectral $k$-means, which correspond to 206.7\%, 238.5\%, and 404.8\% higher MAEs, respectively. More specifically, our proposed pipeline achieves a torso-speed MAE of 0.031~m/s across all test samples. On the 370 samples containing all 3 body part speeds, our model's leg speed MAE is 0.179~m/s, compared with 0.418~m/s for residual-ridge DP, which is a 133.5\% increase in estimation error. Our model also estimates the leg components' relative ridge amplitude with an MAE of $0.093$, while the torso and leg ridge smear widths have an estimation NMSE of $0.016$ and $0.057$, respectively. Our pipeline's ridge parameter estimation performance is further showcased in Table~\ref{tab:varying_psi_synthetic_ridge_params}. On the other hand, the baselines are unable to estimate the spectrogram ridge amplitudes or smear width, as discussed earlier.

\begin{table*}[t]
\centering
\caption{Joint speed-$\psi$ estimation performance on a test set of 500 synthetic walking spectrograms. ``Torso'' denotes the torso-speed metric across all samples; ``Legs'' pools the two leg speeds over the 370 samples in which all three body-part ridges are visible. Speed MAE and RMSE are in m/s.}
\label{tab:varying_psi_synthetic_results}
\footnotesize
\setlength{\tabcolsep}{5.0pt}
\begin{tabular}{lcccccc}
\toprule
& \multicolumn{2}{c}{\textbf{Per-component speed MAE}}
& \multicolumn{3}{c}{\textbf{All speed metrics}}
& \textbf{Geometry factor $\psi$} \\
\cmidrule(lr){2-3}\cmidrule(lr){4-6}\cmidrule(lr){7-7}
\textbf{Method}
& \textbf{Torso} & \textbf{Legs}
& \textbf{MAE} & \textbf{RMSE} & \textbf{NMSE}
& \textbf{$\psi$ MAE} \\
\midrule
Residual-ridge DP
& 0.227 & 0.418 & 0.319 & 0.424 & 0.197 & 0.477 \\
Separated peaks
& 0.257 & 0.455 & 0.352 & 0.469 & 0.232 & 0.477 \\
Weighted spectral $k$-means
& 0.269 & 0.802 & 0.525 & 0.682 & 0.666 & 0.477 \\
\textbf{Proposed}
& \textbf{0.031} & \textbf{0.179}
& \textbf{0.104} & \textbf{0.150} & \textbf{0.027}
& \textbf{0.029} \\
\bottomrule
\end{tabular}
\end{table*}

\begin{table}[t]
\centering
\caption{Ridge amplitude and smear-width estimation performance on the synthetic test set for various walking geometries. ``Torso'' denotes the torso ridge metrics across all samples; ``Leg'' pools the two leg ridge parameters over the 370 samples in which all three body-part ridges are visible. Note that our proposed model can estimate these parameters, while no corresponding baseline estimation method is available.}
\label{tab:varying_psi_synthetic_ridge_params}
\footnotesize
\setlength{\tabcolsep}{5pt}
\begin{tabular}{lc}
\toprule
\textbf{Estimated parameter} & \textbf{NMSE} \\
\midrule
Leg ridge relative amplitude
 & 0.090 \\
Torso ridge smear width
 & 0.016 \\
Leg ridge smear width
 & 0.057 \\
\bottomrule
\end{tabular}
\end{table}

Table~\ref{tab:varying_psi_synthetic_results} further includes a $\psi$ estimation performance evaluation. Varying-$\psi$ estimates $\psi(t)$ with an MAE of 0.029, RMSE of 0.033, and NMSE of 0.001. In comparison, the fixed $\psi=2$ value used has a $\psi$ MAE of 0.477, RMSE of 0.554, and NMSE of 0.497 over the test dataset with various geometries for walks. As expected, the speed estimation performance improvement is correlated with the geometry mismatch and how far the actual walking geometry is from the assumed $\psi = 2$ for the baselines. For walking away/towards the link scenarios, for which $\psi=2$ exactly ($97$ samples), our proposed model and the best-performing baseline achieve average speed MAEs of 0.103 and 0.144~m/s, respectively. Whereas across the two $60^\circ$ diagonal scenarios ($95$ samples), where the baselines have a $\psi$ MAE of 1.106, the corresponding speed errors are 0.125 and 0.543~m/s. In conclusion, our proposed model consistently outperforms the baselines with an increasing advantage for far-from-assumed $\psi$ scenarios. 

\subsection{Joint Speed--Geometry Estimation with Real WiFi}
\label{subsec:results_varying_psi_real}

We finally evaluate joint speed--geometry estimation using four real WiFi walking experiments performed at different path angles relative to the link. A concurrent video collection provides the speed ground truth. The known Tx/Rx locations together with the predefined walking path provide ground-truth $\psi(t)$ for evaluating the network predictions. Table~\ref{tab:varying_psi_real_results} shows the $\psi$ estimation results for each experiment in detail. Moreover, Table~\ref{tab:varying_psi_real_speed_results} compares the performance of the proposed pipeline with that of the baselines.  

\begin{table*}[t]
\centering
\caption{Geometry-factor ($\psi$) estimation performance of our proposed model on four WiFi walking experiments from two subjects across two locations. Related coordinates are expressed in the common coordinate frame, with WiFi TX and RX at $(0,-0.25)$ and $(0,0.25)$ m, respectively.}
\label{tab:varying_psi_real_results}
\footnotesize
\setlength{\tabcolsep}{4.0pt}
\begin{tabular}{llcccccc}
\toprule
\textbf{Experiment No}
& \textbf{Walking path}
& \textbf{Start $(x_0,y_0)$ (m)}
& \textbf{Ref. $\psi$ range}
& \textbf{Pred. $\psi$ range}
& \textbf{$\psi$ MAE}
& \textbf{$\psi$ RMSE}
& \textbf{$\psi$ NMSE} \\
\midrule
1 & Horizontal, $0^\circ$ & $(0.00,1.60)$
& $0.776$--$1.917$ & $0.642$--$1.804$
& $0.209$ & $0.216$ & $0.017$ \\

2 & Diagonal, $-60^\circ$ & $(0.75,2.50)$
& $0.007$--$1.666$ & $0.001$--$1.268$
& $0.333$ & $0.371$ & $0.122$ \\

3 & Vertical, $90^\circ$ & $(2.00,-1.50)$
& $0.001$--$1.757$ & $0.002$--$1.514$
& $0.227$ & $0.245$ & $0.041$ \\

4 & Horizontal, $0^\circ$ & $(1.30,1.60)$
& $1.457$--$1.921$ & $1.438$--$1.921$
& $0.004$ & $0.006$ & $1.1\times10^{-5}$ \\
\midrule
\textbf{Overall} & -- & --
& $0.001$--$1.921$ & $0.001$--$1.921$
& \textbf{$0.193$} & \textbf{$0.209$} & \textbf{$0.045$} \\
\bottomrule
\end{tabular}
\end{table*}

\begin{table}[t]
\centering
\caption{Speed estimation NMSE on the real WiFi walking experiments.}
\label{tab:varying_psi_real_speed_results}
\footnotesize
\setlength{\tabcolsep}{5pt}
\begin{tabular}{lccc}
\toprule
\textbf{Method}
& \textbf{Torso}
& \textbf{Legs}
& \textbf{All speeds} \\
\midrule
Residual-ridge DP
& 0.145 & 0.342 & 0.285 \\
Separated peaks
& 0.183 & 0.346 & 0.295 \\
Weighted spectral $k$-means
& 0.285 & 1.160 & 0.892 \\
\textbf{Proposed}
& \textbf{0.033} & \textbf{0.224} & \textbf{0.166} \\
\bottomrule
\end{tabular}
\end{table}

As can be seen, across all the trials, our proposed model achieves a mean torso-speed MAE of $0.125$~m/s, compared to $0.238$~m/s for residual-ridge DP (the best baseline), which corresponds to an MAE increase of $90.4\%$. Similar trends can be seen for other metrics. In terms of geometry estimation, our pipeline estimates  $\psi(t)$ with an NMSE of $0.045$, significantly reducing it from the baseline value of $0.544$.  

Next, we consider the error in estimating the leg speeds. We note that for angled walking paths, the visibility of external limbs is subject to intermittent occlusion for both video and WiFi sensors, making the reference measurements and speed estimates less reliable. With this limitation in mind, we next report the leg speed estimation performance for the sake of completeness. Our proposed model achieves left and right leg MAEs of $0.332$ and $0.364$~m/s, respectively. Hence, the best-performing baseline, residual-ridge DP, with leg speed MAEs of $0.446$ and $0.449$, increases the estimation MAE by $28.6\%$ when averaged over two legs. 

Finally, we evaluate the joint model using an existing real-world dataset. Because no available datasets contain walking trajectories with time-varying $\psi$, we use a dataset~\cite{cai2020teaching,gym_dataset} of 26 walks from 13 subjects in which $\psi \approx 2$. Without being provided with this geometry factor, the model jointly estimates $\psi(t)$ and all three speed components, achieving an all-speed NMSE of $0.093$, and a $\psi(t)$ NMSE of $0.005$.

Overall, our extensive evaluation demonstrates the strength and generalizability of the proposed pipeline across two complementary settings. When the geometry is known, the fixed-$\psi$ model accurately disentangles multiple overlapping speed components and generalizes well from purely synthetic training to real-world WiFi measurements. In the more challenging changing-$\psi$ scenarios, the joint model successfully recovers distinct speed components without prior knowledge of $\psi$, while also accurately estimating the underlying geometry factor. Together, these results establish the pipeline’s ability to recover fine-grained motion and geometry from complex WiFi spectrograms.

\subsection{Computation Time}
\label{subsec:computation_time}

Using an Intel Core i7-9700K CPU and an NVIDIA GeForce RTX 2080 GPU, we report the average computation time over 10 runs of our proposed pipeline. The raw CSI preprocessing took $18.17 \pm 0.35$ ms per second of input, whereas the inference took $4.79 \pm 1.4$ ms per second of input. Our overall pipeline required 22.96 ms per second of input (real-time factor, RTF = 0.023), which corresponds to 43.54 times faster than real-time processing. In comparison, three baselines required $88.54 \pm 1.51$ ms per second of input for inference.

\section{Conclusions}
\label{sec:conclusion}
In this paper, we establish a new foundation that can effectively disentangle reflector speed from geometry, jointly recovering the speed, geometry factor, relative amplitude, and width of each dominant Doppler ridge. More specifically, we first developed a parametric representation of WiFi spectrograms and methodically established its low-dimensionality through a systematic analysis of a large and diverse human-activity dataset. Building on this compact representation, we then designed a physics-informed autoencoder whose structured bottleneck and differentiable RF forward model enforce physically meaningful estimates of reflector speed and geometry. We further eliminated the need for real WiFi training data by introducing a synthetic-to-real training framework. We extensively validated the proposed framework using both independently generated synthetic test sets as well as 31 real WiFi experiments. The results demonstrated that the proposed pipeline can robustly recover the underlying geometry, speeds, Doppler-ridge amplitudes, and ridge widths across all settings, while substantially outperforming the strongest baselines.

\section*{References}
\bibliography{main}
\bibliographystyle{IEEEtran}

\end{document}